\documentclass[draft,11pt]{article}
\usepackage{amssymb}
\newcommand{\showlabel}[1]{}
\newcommand{\beq}[1]{\showlabel{#1}\begin{equation}\label{#1}}
\newcommand{\eeq}{\end{equation}}
\newcommand{\req}[1]{(\ref{#1})}

\newcommand{\eop}{\rule{0.7em}{0.7em}}
\newcommand{\eq}{\triangleq}
\renewcommand{\emptyset}{\varnothing}
\newcommand{\str}[1]{\left(\strut{#1}\right)}
\newcommand{\strf}[1]{\left\{\strut{#1}\right\}}
\newcommand{\strk}[1]{\left[\strut{#1}\right]}
\renewcommand{\r}{{\bf r}}
\newcommand{\y}{{\bf y}}
\newcommand{\x}{{\bf x}}
\newcommand{\p}{{\sf p}}
\newcommand{\I}{{\rm I}}
\renewcommand{\l}{\ell}
\renewcommand{\i}{\imath}

\newcommand{\0}{{\bf 0}}
\newcommand{\V}{{\rm V}}

\begin{document}

 Published in : {\em Proceedings of the 6th International Workshop on Model Oriented Design and Analysis}, 
 pp.~63-75, Physica-Verlag Heidelberg, Puchberg/Schneeberg, Austria, 2001.

\begin{flushright}
{\bf A.G.\,D'yachkov\footnotemark[1],
A.J.\,Macula\footnotemark[2],
D.C.\,Torney\footnotemark[3],
P.A.\,Vilenkin\footnotemark[1]}
\end{flushright}
\footnotetext[1]{Department of Probability Theory,
Faculty of Mechanics \& Mathematics,
Moscow State University, Moscow, 119899, Russia
(e-mail: dyachkov@artist.math.msu.su,
paul@vilenkin.dnttm.ru).
The work of A.G.~Dyachkov and P.A.~Vilenkin is supported
by the Russian Foundation of Basic Research, qrant
98-01-241.} \footnotetext[2]{Department of Mathematics,
State University New York,
College at Geneseo, Geneseo, NY 14454 USA
(e-mail: macula@geneseo.edu).}
\footnotetext[3]{Los Alamos National Laboratory,
MS K710, Los Alamos, NM 87545 USA
(e-mail: dct@ipmati.lanl.gov).}
\par
\vspace*{1cm}
\par
\begin{center}
{\Large\bf
Two Models of Nonadaptive Group Testing for\\
Designing Screening Experiments}
\end{center}

\begin{abstract}
We discuss two non-standard models of nonadaptive combinatorial
search which develop the conventional disjunct search model~\cite{dh93}
for a small number of defective elements contained in a finite ground set or
a population.  The first model called a {\it search of defective
supersets} ({\it complexes\/}) was suggested in~\cite{dmtvy00,dmtv00}. The
second model which can be called a {\it search of defective subsets in
the presence of inhibitors} was introduced for the case of an
adaptive search in~\cite{fkkm97,dv98}. For these models, we
study the constructive search methods based on the known
constructions for the disjunct model from~\cite{ks64,dmr00,dmr00_2}.
\end{abstract}

\section{Description of the Models}

We use symbol $\eq$ to denote definitional equalities.

The standard {\em disjunct} search model for {\em designing screening
experiments} (DSE)~\cite{d97} has the following form.
Let there be a {\it population} containing $t$ distinguishable {\it samples}.
We identify it with the set $[t]\eq\{1,2,\ldots,t\}$. Assume that
the population includes an unknown {\it defective} subset $\p\subset[t]$.
We call elements of $\p$ {\it defective samples}.

Our aim is to detect $\p$ using a number of {\it group tests}.
Each group test is defined by a subset ({\it testing group} or {\it pool\/})
$G\subset[t]$. The result $r(G,\p)$ of this test assumes a binary
value 0 ({\it negative\/}) or 1 ({\it positive\/}) according to the
following rule:
\beq{res}
r(G,\p)\eq\left\{\begin{array}{ll}
1,&G_n\cap\p\ne\emptyset,\\[5pt]
0,&G_n\cap\p=\emptyset.
\end{array}\right.
\eeq
One can see that a group test detects whether the testing group intersects
with the defective subset or not.

In the {\it adaptive} disjunct model each group test $G_{n+1}$
is choosen according to the results of the previous tests $G_1,\ldots,G_n$.
This model is considered in \cite{dh93}.

We consider the {\em nonadaptive} disjunct model,
in which all pools must be constructed before any test is performed.
Thus, an explorer is not allowed to use the results of the
previous tests to construct the next ones. Such model usually
occurs in the problems of molecular biology \cite{fkkm97,dmr00_2}
when defferent tests can be performed simultaneously
as one experiment (one step). Each experiment is expensive, so we are
interested in detecting $\p$ using only one step, i.e., exactly
the nonadaptive search strategy.

We also consider the restriction on the number of defective samples:
$|\p|\le s$, where $s$ is the given positive integer, $s<t$.
Such condition is usually used in the group testing problems. In many
cases when such condition does not occur naturally in problems under
consideration it can be justified by the probabilistic arguments.
If any sample can be defective with some small probability, then
one can choose $s$ so that the inequality $|\p|\le s$ holds with
high probability.

Note, that for the practical applications of this model we assume that
$s\ll t$. If this condition does not hold, then the present model
is not suitable.

{\bf Definition 1.}
A series of $N$ nonadaptive tests $X\eq(G_1,G_2,\dots,G_N)$ which allows
to identify any defective subset $\p\subset[t]$, $|\p|\le s$,
is called a {\em disjunct $s$-design}~\cite{dh93} of {\em length} $N$
and {\em size}~$t$. Note that we suppose no error in the results
of these test.

In general case to detect a defective subset $\p$ using a
disjunct $s$-design one needs to check all possible subsets.
Below we introduce an important class of designs called {\it superimposed
codes} for which the decoding algorithm is much more simple.

\medskip
This model can be generalized in several ways. We can consider the
{\it list decoding procedure} in which one should construct a subset
$\p'\subset[t]$ such that $\p\subseteq\p'$ and the size of
$\p'\backslash\p$ is not very large. This model was considered in
\cite{dr83} and some other papers. We also can consider designs
which correct errors in the results \cite{drr89}.

\medskip
In the present paper we introduce two generalizations
of the disjunct model of DSE.

\begin{enumerate}
\item {\bf Nonadaptive search of defective supersets.}

Let there be a population $[t]=\{1,2,\ldots,t\}$.
Assume that there exists an unknown {\it superset} ({\em complex\/}) $\p$
which is composed of a number of subsets $P\subset[t]$:
\beq{superset}
\p\eq\{P_1,P_2,\dots,P_k\},\quad P_i\subset[t],\qquad
P_i\not\subset P_j\mbox{ for }i\ne j.
\eeq
To detect $\p$ one can use a number of group tests $G\subset[t]$ for which
the result $r(G,\p)$ is positive if and only if $G$ includes at least
one subset $P\subset[t]$ being a member of the complex $\p$, and
negative otherwise:
\beq{res1}
r(G,\p)\eq\left\{\begin{array}{ll}
1,&\exists P\in\p\,:\,P\subseteq G,\\[5pt]
0,&\mbox{otherwise}.
\end{array}\right.
\eeq

We consider the following restrictions on complexes $\p$: the number of
elements $P\in\p$ does not exceed $s$ and the size $|P|\le\l$ for any
$P\in\p$, where $s$ and $\l$ are the given positive integers,
$s+\l\le t$.

{\bf Definition 2.}
A series of $N$ nonadaptive tests $X=(G_1,G_2,\dots,G_N)$ which allows
to identify any such complex $\p$ is called a {\em superset $(s,\l)$-design}
of length $N$ and size $t$.

Obviously, for $\l=1$ this model is identical to the conventional
disjunct model, because each subset $P\in\p$ is composed of exactly
one element.

One can easily understand the necessity of the additional
condition in \req{superset}: if $P_i\subset P_j$, then we cannot detect
the defective property of $P_j$.

\medskip
Consider the following simple example of such application. We have some
chemical (or medical) substances. We assume that some combinations
of them may be dangerous. To detect these combinations we can perform
group tests, i.e., put some of the samples together and test whether
the obtained substance is dangerous. The result is positive if the group
contains one or more dangerous combinations, that leads to the
given model.

\newpage
\item {\bf Nonadaptive search of
defective subsets in the presence of inhibitors.}

Return to the base disjunct model and assume that along with defective
subset $\p\subset[t]$ there exists a subset $\I\subset[t]$,
$\p\cap \I=\emptyset$. We call samples from this set {\em inhibitors}.

Inhibitors make the result of a group test negative despite the
existence of defective elements in testing group $G\subset[t]$:
\beq{res2}
r(G,\p,\I)\eq\left\{\begin{array}{ll}
1,&G\cap\p\ne\emptyset\quad\mbox{and}\quad
G\cap \I=\emptyset,\\[5pt]
0,&\mbox{otherwise}.
\end{array}\right.
\eeq

{\bf Definition 3.}
A series of $N$ nonadaptive tests $X=(G_1,G_2,\dots,G_N)$ which allows
to identify any defective subset $\p$, $|\p|\le s$, in the presence of not
more then $\i$ inhibitors is called an {\em inhibitory $(s,\i)$-design}
of length $N$ and size $t$, where $s\ge1$ and $\i\ge0$ are the given
integers, $s+\i\le t$.

Obviously for $\i=0$ this model is identical to the
conventional disjunct model of DSE.

\medskip
The current model arises in applications in which some samples in a
population are not defective, but affect the testing result, namely,
make it always negative. If a test has a form of some chemical reaction,
then it may stop due to the inhibitor. This illustration gives the name
to the current model.
\end{enumerate}

\medskip
The rest part of the present paper is organized as follows.
In section~2 we consider important classes of designs for
the models under consideration. The detecting algorithm for
these types of designs are simple. In section 3 we consider
a constructive method for such designs. Section~4 contains
several examples.

\section{Superimposed codes}

\subsection{Notations}

Let $[t]$ be a population of samples. Consider an arbitrary series of $N$
testing groups $X=(G_1,\ldots,G_N)$, $G_n\subset[t]$ for $n=1,2,\ldots,N$.
Following~\cite{dh93}, we encode it by the binary
$N\times t$ {\em incidence matrix} $X=\|x_n(u)\|$, where index
$n=1,2,\dots N$ denotes a row number and index $u=1,2,\dots t$ denotes
a column number. An element $x_n(u)$ of this matrix has the form
$$
x_n(u)\eq\cases{1, & if $u\in G_n$,\cr
                0, & otherwise.\cr}
$$
The $n$-th row $\x_n\eq\str{x_n(1),x_n(2),\dots,x_n(t)}$ of matrix $X$
encodes the $n$-th test $G_n$. The $u$-th column
$\x(u)\eq\str{x_1(u),x_2(u),\ldots,x_N(u)}$ is called the $u$-th
{\it codeword}.

Denote by $\x\bigvee\y$ $(\x\bigwedge\y)$ the componentwise
disjunction (conjunction) of binary vectors $\x$ and $\y$
of the same length. We say that a vector $\x$ {\em covers}
a vector $\y$ if $\x\bigvee\y=\x$. For a matrix $X$ and subset
of columns $\tau\subset[t]$ consider the disjunction and conjunction
$$
\V(X,\tau)\eq\bigvee_{u\in\tau}\x(u),\qquad
\Lambda(X,\tau)\eq\bigwedge_{u\in\tau}\x(u).
$$
If $\tau=\emptyset$ then we put $\V(X,\emptyset)\eq\0=(0,0,\ldots,0)$.

\subsection{Disjunct model of DSE}

Let $\p\subset[t]$ be a defective subset. Denote by $\r(X,\p)$ the
binary vector of results of all $N$ tests:
$\r(X,\p)\eq\str{r(G_1,\p),r(G_2,\p),\ldots,r(G_N,\p)}$.
From definition \req{res} one can see that it has the form
of the disjunction of codewords:
\beq{rr1}
\r(X,\p)=\V(X,\p)=\bigvee_{u\in\p}\x(u).
\eeq

{\bf Definition 1'.} $X$ is a disjunct $s$-design iff for any two different
subsets $\p_1,\p_2\subset[t]$, $|\p_1|\le s$, $|\p_2|\le s$,
the result vectors $\r(X,\p_1)$ and $\r(X,\p_2)$ are different.

\medskip
{\bf Definition 4} \cite{ks64}. Let $s$ be an integer, $0<s<t$. A binary
$N\times t$ matrix $X$ is called a {\it superimposed $s$-code}
of length $N$ and size $t$ if for any subset $\p\subset[t]$,
$|\p|\le s$, and any sample $u\in[t]\backslash\p$ the codeword
$\x(u)$ is not covered by the disjunction $\r(X,\p)$ \req{rr1}.
One can see that it is equivalent to the following condition:
for the given pair $(\p,u)$ there exists a row number $n\in[N]$
such that $x_n(u)=1$ and $x_n(u')=0$ for all $u'\in\p$.

\medskip
{\bf Lemma 1} \cite{ks64}.
{\it Any superimposed $s$-code is a disjunct $s$-design.}

{\bf Proof.} Obviously, for any sample $u\in\p$ the disjunction
$\r(X,\p)$ \req{rr1} covers the codeword $\x(u)$. If $X$ is a
superimposed $s$-code and $|\p|\le s$, then from definition 4 it
follows that for any sample $u\notin\p$ this disjunction does
not cover the codeword $\x(u)$. Thus, one can easily detect $\p$
and $X$ is a disjunct $s$-design.~\eop

\medskip
If $X$ is a disjunct $s$-design, then the complexity of the trivial
algorithm of detecting $\p$ is $\sim{t\choose s}$ because we need
to perform an exhaustive search over all subsets $\p\subset[t]$,
$|\p|\le s$. The superimposed $s$-code condition provides the simple
decoding algorithm: given the vector $\r(X,\p)$ and any sample
$u\in[t]$ one can easily detect whether $u$ is defective or not.
The complexity of this algorithm is $\sim t$.

Superimposed $s$-codes were introduced in \cite{ks64} and studied in
many papers \cite{dr83,drr89,d97,dmr00,dmr00_2}. Below we consider the
similar types of designs for the models of searching supersets and
searching subsets in the presence of inhibitors.

\subsection{Search of defective supersets}

Let $X$ be a binary $N\times t$ matrix which encodes a search
strategy as described before. Let $\p$ be a defective superset
(complex) composed of a number of subsets $P\subset[t]$ \req{superset}.
Using the definition \req{res1} one can easily prove that the
result vector $\r(X,\p)\eq\str{r(G_1,\p),r(G_2,\p),\ldots,r(G_N,\p)}$
in this model has the form
\beq{rr2}
\r(X,\p)=\bigvee_{P\in\p}\Lambda(X,P)=
\bigvee_{P\in\p}\bigwedge_{u\in P}\x(u).
\eeq

{\bf Definition 2'.} $X$ is a superset $(s,\l)$-design (see Def. 2) iff
for any two different complexes $\p_1$ and $\p_2$ composed of
not more then $s$ subsets which sizes do not exceed $\l$
the results $\r(X,\p_1)\ne\r(X,\p_2)$.

\medskip
{\bf Definition 5} \cite{mp88}.
Let $s$ and $\l$ be positive integers, $s+\l\le t$.
A binary $N\times t$ matrix $X$ is called a {\it superimposed $(s,\l)$-code}
if for any subsets $S,L\subset[t]$, $|S|\le s$, $|L|\le\l$ and
$S\cap L=\emptyset$ there exists a row number $n\in[N]$ for which
$x_n(u)=1$ for any $u\in L$ and $x_n(u')=0$ for any $u'\in S$.

\newpage
One can see that for $\l=1$ this definition coincides with Def. 4
of superimposed $s$-codes. Superimposed $(s,\l)$-codes were first
introduced in \cite{mp88} for criptography applications. It the present
paper we study them for the search problems. Below we introduce some
simple results which are going to be published in a more detailed form
in \cite{dmtv00}.

\medskip
{\bf Lemma 2.} {\it Any superimposed $(s,\l)$-code is a superset
$(s,\l)$-design.}

{\bf Proof.} Let $\p$ be a superset \req{superset}, where $k\le s$
and $|P_i|\le\l$ for any $i$. Let $X$ be a superimposed $(s,\l)$-code.
Our aim is to detect $\p$ given the vector $\r(X,\p)$ \req{rr2}.

Any subset $P\subset[t]$, $|P|\le\l$, belongs to one of the
following two types:
\begin{itemize}
\item[$(\alpha)$] there exists $P_i\in\p$ such that $P_i\subseteq P$;
\item[$(\beta)$] $P$ does not contain any defective subset $P_i\in\p$.
\end{itemize}
Let us show that for any subset $P$ we can detect whether it
belongs to the class $(\alpha)$ or $(\beta)$ given the result $\r(X,\p)$.
Indeed, if $P$ satisfies $(\alpha)$, then from \req{rr2} it follows
that the vector $\r(X,\p)$ covers the conjunction $\Lambda(X,P)$.

Assume that $P$ satisfies condition $(\beta)$. Then all $k$ sets
$P_i\backslash P$ for $1\le i\le k$ are not empty. Take a sample
from each of these sets and construct a set $S\subset[t]$ containing
all these samples, $|S|\le k\le s$. Put $L\eq P$, $|L|\le\l$. For
this pair of sets $(S,L)$ take a row number $n\in[N]$ according to
definition 5. One can easily prove that $n$-th components of
all vectors $\Lambda(X,P_i)$, $1\le i\le k$, are zeros, and thus
the $n$-th component of $\r(X,\p)$ \req{rr2} is zero. And the $n$-th
component of the vector $\Lambda(X,P)$ is 1. This proves that
for the case $(\beta)$ the vector $\r(X,\p)$ does not cover $\Lambda(X,P)$.

The class $(\alpha)$ contains both defective subsets $P_i\in\p$
and subsets $P$ such that $P_i\varsubsetneq P$ for some $i$. To
separate these cases note that if $P\in\p$, then all subsets of $P$
satisfy $(\beta)$. And for the second case there exists a subset
of $P$ which satisfies $(\alpha)$.

We showed that given the vector $\r(X,\p)$ it is possible to
detect $\p$, that proves the statement. Moreover, we obtained the
algorithm of such detection. Given a subset $P\subset[t]$ this
algorithm detects whether $P\in\p$ or not.~\eop

\medskip
If $X$ is a superset $(s,\l)$-design, then the complexity of the trivial
algorithm of detecting a complex $\p$ is $\sim{{t\choose\l}\choose s}$
because we need to perform an exhaustive search over all supersets $\p$.
The superimposed $(s,\l)$-code condition provides the simple
decoding algorithm: given the vector $\r(X,\p)$ and any subset $P\subset[t]$,
$|P|\le\l$, one can easily detect whether $P\in\p$ or not.
The complexity of this algorithm is $\sim {t\choose \l}$.

\subsection{Search of defective subsets in the presence of inhibitors}

Let $X$ be a binary $N\times t$ matrix which encodes a search
strategy as described before. Let $s\ge1$ and $\i\ge0$ be integers,
$s+\i\le t$. Denote by $\pi(t,s,\i)$ the set of all
possible pairs $(\p,\I)$, where $\p$ is a defective set and
$\I$ is a set of inhibitors:
$$
\pi(s,\i,t)\eq\strf{(\p,\I)\,:\,\p,\I\subset[t],\,
1\le|\p|\le s,\,0\le|\p|\le\i,\,\p\cap \I=\emptyset}.
$$
For a pair of binary symbols $x,y\in\{0,1\}$ we define the
{\it inhibition of $x$ due to $y$} operation:
$$
x\setminus y\eq\cases{1, & if
$x=1$ and $y=0$,\cr
0, & otherwise.\cr}
$$
For a pair of binary vectors $\x$ and $\y$ we denote by $\x\setminus\y$
the componentwise inhibition operation. Note that $\x\setminus\y=\0$
iff $\x$ is covered by $\y$, and
$\x\setminus\y=\x$ iff the conjunction $\x\bigwedge\y=\0$.

Let $(\p,\I)\in\pi(s,\i,t)$ be a pair of defective set and inhibitor set.
Using the definition \req{res2} one can easily prove that the
result vector $\r(X,\p,\I)\eq\str{r(G_1,\p,\I),r(G_2,\p,\I),\ldots,r(G_N,\p,\I)}$
in this model has the form
\beq{rr3}
\r(X,\p)=\V(\p)\backslash\V(\I)=\left.\bigvee_{u\in\p}\x(u)\right\backslash
\V(\I)=\bigvee_{u\in\p}\strk{\x(u)\backslash\V(\I)}.
\eeq

{\bf Definition 3'.} $X$ is an inhibitory $(s,\i)$-design (see Def. 3) iff
for any two pairs $(\p,\I),(\p',\I')\in\pi(t,s,\i)$, such that $\p\ne\p'$,
the result vectors $\r(X,\p,\I)\ne\r(X,\p',\I')$. Note that this condition
allows to detect the defective subset $\p$ but not the inhibitory set $\I$.

\medskip
{\bf Definition 6.} Let $s\ge1$ and $\i\ge0$ be integers, $s+\i\le t$.
A binary $N\times t$ matrix $X$ is called an {\it inhibitory $(s,\i)$-code}
if it is a superimposed $(s+\i)$-code (see Def.~4).

\medskip
{\bf Lemma 3.} {\it Any inhibitory $(s,\i)$-code is an inhibitory
$(s,\i)$-design.}

{\bf Proof.} Let $(\p,\I)\in\pi(t,s,\i)$ be a pair of defective and inhibitory
sets and $X$ be an inhibitory $(s,\i)$-code, i.e., a superimposed
$(s+\i)$-code. We should detect $\p$ given the vector $\r(X,\p,\I)$ \req{rr3}.

Let us call a sample $u\in[t]$ {\it $\i$-acceptable due to the vector
$\r(X,\p,\I)$} if there exists a subset $\I'\subset[t]$ such that
$u\notin \I'$, $|\I'|\le\i$ and the vector $\r(X,\p,\I)$ covers
$\x(u)\setminus\V(\I')$.

If $u\in\p$, then $u$ is acceptable because all conditions hold for
$\I'=\I$. Assume that $u\notin\p$. Then for any subset $\I'\subset[t]$,
$u\notin \I'$, $|\I'|\le\i$, consider a pair $(\p\cup \I',u)$ and take
a row number $n\in[N]$ according to the definition of superimposed
$(s+\i)$-code (Def.~4). One can easily prove that the $n$-th component
of the vector $\r(X,\p,\I)$ is zero and the $n$-th component of
$\x(u)\setminus\V(\I')$ is 1. So the first vector does not cover the
second one. Since this is true for any $\I'$, the sample $u$ is
not acceptable.

We proved that a sample is defective iff it is $\i$-acceptable due
to $\r(X,\p,\I)$. Obviously, one can check this given only the result
$\r(X,\p,\I)$. This completes the proof of the statement and also
gives the decoding algorithm. Note that one can consider only
such subsets $\I'$ for which $\V(\I')\bigwedge\r(X,\p,\I)=\0$.~\eop

\medskip
The complexity of the trivial algorithm of detecting a subset $\p$
for an arbitrary inhibitory $(s,\i)$-design is
$\sim{t\choose s+\i}\cdot{s+\i\choose s}$ because we should perform
an exhaustive search over all pairs from the set $\pi(t,s,\i)$.
The inhibitory $(s,\i)$-code condition provides the simple
decoding algorithm: given the vector $\r(X,\p)$ and a sample $u\in[t]$
we need to check all subsets $\I'$.
The complexity of this algorithm is $\sim t\cdot{t\choose\i}$.

\section{Concatenated construction for superimposed codes}

In the present section we consider a concatenated construction
for superimposed $(s,\l)$-codes. For special case $\l=1$ this leads to
the construction for superimposed $s$-codes. For this case the
similar method was suggested in \cite{ks64} and developed in
\cite{dmr00,dmr00_2}.

\medskip
{\bf Definition 7} \cite{fgu69}.
Let $q\ge2$ be an integer and $X=\|x_n(u)\|$ be an
$N\times t$ $q$-ary matrix: $n\in[N]$, $u\in[t]$, $x_n(u)\in[q]$.
Let $s$ and $\l$ be positive integers, $s+\l\le t$. Then $X$ is
called a {\it $q$-ary separating $(s,\l)$-code} if for any two
subsets $S,L\subset[t]$, $|S|\le s$, $|L|\le\l$, $S\cap L=\emptyset$,
there exists a row number $n\in[N]$ such that the corresponding
coordinate sets $S_n$ and $L_n$ do not intersect, where
$$
L_n\eq\strf{x_n(u)\,\,:\,\,u\in L}\subset[q],\qquad
S_n\eq\strf{x_n(u')\,\,:\,\,u'\in S}\subset[q].
$$

Note that for $q=2$ definition 7 does not coincide with definition
5 of binary superimposed $(s,\l)$-code. Binary separating $(2,2)$-codes
were studied before for certain applications \cite{fk84,s94}.

\medskip
{\bf Lemma 4.} {\it Let $X^{(q)}$ be a $q$-ary separating $(s,\l)$-code
of size $t^{(q)}$ and length $N^{(q)}$. Let $X'$ be a binary
superimposed $(s,\l)$-code of size $q$ and length $N'$. Then there
exists a binary superimposed $(s,\l)$-code $X$ of size $t=t^{(q)}$
and length $N=N^{(q)}\cdot N'$.}

{\bf Proof.}
Consider the code $X$ obtained by the concatenation of
codes $X^{(q)}$ and $X'$, i.e., each $q$-ary symbol
$\theta\in[q]$ in matrix $X^{(q)}$ is replaced by the
$\theta$-th codeword from $X'$:
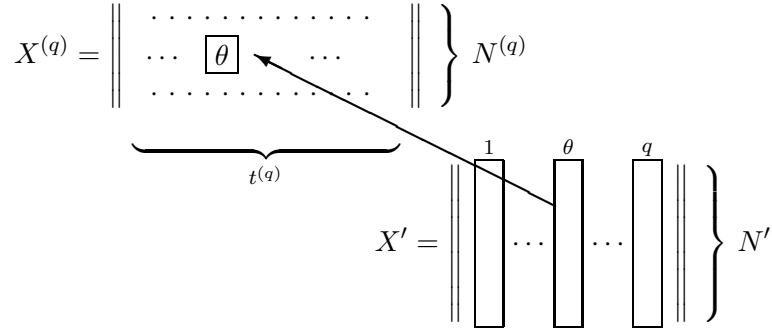
\begin{center}
\unitlength=1mm
\begin{picture}(125,40)
\newcommand{\mat}{{\begin{array}{ccc}
\multicolumn{3}{c}{\hbox to 3.2cm{\leaders\hbox{$\,\cdot\,$}\hfill}}\\
\cdots&\fbox{$\theta$}&\cdots\\
\multicolumn{3}{c}{\hbox to 3.2cm{\leaders\hbox{$\,\cdot\,$}\hfill}}
\end{array}}}
\put(10,33){$\displaystyle X^{(q)}=\left\|\vphantom{\mat}\right.
\lefteqn{\raisebox{-0.4cm}{$\underbrace{\phantom{\mat}}_{t^{(q)}}$} }
\mat\left.\vphantom{\mat}\right\|\,
\left.\vphantom{\mat}\right\}\,N^{(q)}$}
\newcommand{\clm}{{\fbox{\parbox[c][2cm]{0.15cm}{\hfil}}}}
\put(58,8){$\displaystyle X'=\left\|\vphantom\clm\right.
\stackrel{1}{\clm}\cdots\stackrel{\theta}{\clm}\cdots\stackrel{q}{\clm}
\left.\vphantom\clm\right\|\,\left.\vphantom\clm\right\}\,N'$}
\thicklines
\put(82,14){\vector(-2,1){40}}
\end{picture}
\end{center}
One can easily prove the superimposed $(s,\l)$-code property for $X$.~\eop

\medskip
A $q$-ary code $X^{(q)}$ is called an {\it external code}, and
a binary code $X'$ is called an {\it internal code}. To construct
concatenated codes with large sizes we need $q$-ary external codes with
large sizes and binary internal codes with small sizes. Below we discuss
some simple methods for constructing them.

\medskip
{\bf Lemma 5} (trivial code). {\it For any positive integers
$s$, $\l$ and $t$, $s+\l\le t$, there exists a superimposed
$(s,\l)$-code of size $t$ and length}
$$
N=\min\strf{{t\choose s};{t\choose\l}}.
$$

{\bf Proof.} To obtain a code of length $N_1={t\choose s}$ take the
binary $N_1\times t$ matrix which rows are all possible binary
vectors of length $t$ having exactly $s$ zeros. To obtain a code
of length $N_2={t\choose\l}$ take the binary $N_2\times t$ matrix
which rows are all possible binary vectors of length $t$ having
exactly $\l$ ones. Obviously, both these matrixes satisfy the
$(s,\l)$-code property.~\eop

\medskip
This trivial construction allow to construct superimposed codes
for all possible values of $s$, $\l$ and $t$. But it is reasonable
to use it only for small values of $t$. Note that for $t=s+\l$
the trivial code is optimal (has the smallest possible length).

Several methods exist for constructing small superimposed codes
for some special values of $s$ and $\l$. Some tables of codes
for $\l=1$ can be found in \cite{dmr00,dmr00_2}. The values $s=\l=2$
are considered in \cite{s94,dmtvy00,dmtv00} and below.

\medskip
Finally we discuss a method of constructing large $q$-ary separating
codes to be used in concatenated construction. It is based on MDS-codes.
Since these codes are well-known we do not consider their properties
in details.

\medskip
{\bf Definition 8} \cite{ms83}.
Any $q$-ary code of size $t=q^k$, length $n$ and the
Hamming distance $d=n-k+1$ is called a {\em maximal
distance separable} code (MDS-code) with parameters $(q,k,n)$.

\medskip
{\bf Lemma 6} \cite{dmtvy00,dmtv00}.
{\it If $n\ge s\l(k-1)+1$ and $q^k\ge s+\l$, then any
MDS--code with parameters $(q,k,n)$ is a $q$-ary
separating $(s,\l)$-code.}

\medskip
{\bf Lemma 7} \cite{ms83}.
{\em For any positive integer $\lambda$ and any prime power
$q\ge\lambda$ there exists an MDS--code with parameters
$(q,\lambda+1,q+1)$ called the Reed--Solomon code}.

\medskip
Using Reed--Solomon code for the concatenation
construction, we obtain

{\bf Lemma 8} \cite{dmtvy00,dmtv00}. {\it Let $s$, $\l$,
$\lambda$ be positive integers, and $q\ge s\l\lambda$ be a
prime power. Assume that there exists a binary superimposed
$(s,\l)$-code of size $q$ and length $N_1$. Then there exists
a binary superimposed $(s,\l)$-code of size $t=q^{\lambda+1}$
and length $N=N_1(s\l\lambda+1)$.}

\section{Examples}

{\bf 1.} The best known superimposed $2$-code of size $t=12$
has length $N=9$:
\beq{9x12}
X=\pmatrix{0&0&1&1&1&1&0&0&0&0&0&0\cr
           0&0&1&0&0&0&1&1&1&0&0&0\cr
           0&0&1&0&0&0&0&0&0&1&1&1\cr
           0&1&0&1&0&0&1&0&0&1&0&0\cr
           0&1&0&0&1&0&0&1&0&0&1&0\cr
           0&1&0&0&0&1&0&0&1&0&0&1\cr
           1&0&0&1&0&0&0&0&1&0&1&0\cr
           1&0&0&0&1&0&1&0&0&0&0&1\cr
           1&0&0&0&0&1&0&1&0&1&0&0\cr}.
\eeq
It is the first known code for which $N<t$. For all sizes $t_1<t$
the smallest known codes are trivial. Note that the trivial
superimposed $s$-code of size $t$ is the identity $t\times t$ matrix.

Note also that the matrix \req{9x12} is an inhibitory $(1,1)$-code,
see Def.~6.

\medskip
{\bf 2.} Some examples of small superimposed $(2,2)$-codes
are known. For $t=4$, the optimal code is trivial and has length
$N={4\choose 2}=6$. For $t=5$, the optimal code is also trivial and
has length $N={5\choose 2}=10$. For $t=6$, $t=7$ and $t=8$
the optimal code has length $N=14$. It can be obtained by the
concatenated method from the following $3\times 8$ quaternary matrix:
$$
C^{(4)}=\left(
\begin{array}{cccccccc}
4&2&3&1&2&4&1&3\\
2&4&1&3&2&4&1&3\\
1&1&2&2&3&3&4&4
\end{array}\right),
$$
which is a separating $(2,2)$-code. It can be concatenated with
the trivial superimposed $(2,2)$-code of size $4$ and length $6$.
This leads to the superimposed $(2,2)$-code of size $t=8$ and length $N=18$.
Examining this code, one can see, that there are two rows in it,
which are repeated three times.  Removing the copies, we obtain the binary
superimposed $(2,2)$-code of length $N=14$.

The following table gives several numerical values of the known
superimposed $(2,2)$-codes. Some of them were obtained with the
help of V.\,S.\,Lebedev.
\vspace{0.3cm}
\begin{center}
\begin{tabular}{rcccccccccccccc}
\hline
$t=\quad$ & 4 & 5 & 8 & 12 & 16 & 20 & 25 & 64
& 121 & 512  & 1331 & $2^{12}$ & $2^{16}$ & $2^{20}$\\
$N\le\quad$ & 6 & 10 & 14 & 22 & 28 & 38 & 50 & 70
& 110 & 126 & 198 & 252 & 364 & 476\\
\hline
\end{tabular}
\end{center}

Some of these codes were known before \cite{s94}. Most of them are new.
In the paper \cite{dmtv00} an improved table of codes is going to be
published.

\end{document}